# DeepDGA: Adversarially-Tuned Domain Generation and Detection

Hyrum S. Anderson, Jonathan Woodbridge, and Bobby Filar

{hyrum,jwoodbridge,bfilar}@endgame.com
Endgame, Inc.

*Abstract*—Many malware families utilize domain generation algorithms (DGAs) to establish command and control (C&C) connections. While there are many methods to pseudorandomly generate domains, we focus in this paper on detecting (and generating) domains on a per-domain basis which provides a simple and flexible means to detect known DGA families. Recent machine learning approaches to DGA detection have been successful on fairly simplistic DGAs, many of which produce names of fixed length. However, models trained on limited datasets are somewhat blind to new DGA variants.

In this paper, we leverage the concept of generative adversarial networks to construct a deep learning based DGA that is designed to intentionally bypass a deep learning based detector. In a series of adversarial rounds, the generator learns to generate domain names that are increasingly more difficult to detect. In turn, a detector model updates its parameters to compensate for the adversarially generated domains. We test the hypothesis of whether adversarially generated domains may be used to augment training sets in order to harden other machine learning models against yet-to-be-observed DGAs. We detail solutions to several challenges in training this character-based generative adversarial network (GAN). In particular, our deep learning architecture begins as a domain name auto-encoder (encoder + decoder) trained on domains in the Alexa one million. Then the encoder and decoder are reassembled competitively in a generative adversarial network (detector + generator), with novel neural architectures and training strategies to improve convergence.

Results show that domains generated from a GAN to bypass the GAN's detector also bypass a random forest classifier that leverages hand-crafted features. Conversely, by augmenting the training set with these adversarial examples, the random forest classifier is able to detect with greater efficacy DGA malware families not seen during training.

## I. INTRODUCTION

Like any defensive technology, machine learning models are subject to false positives and false negatives. An important step in delivering a model as part of a product consists of assessing and (if possible) patching model vulnerabilities (e.g., certain malware families for a malware detector, etc.). While this expert-guided process may always exist, we test a key thesis of [1]: *adversarial examples*[1]—artificial samples which a machine learning model misidentifies—can be discovered automatically and used to augment a training dataset to harden (i.e., make more robust) machine learning models. While this thesis has been confirmed generally [2], [1], we propose a novel framework for generating these adversarial examples using a generative adversarial network for a natural language-based domain generation algorithm (DGA) detector.

DGAs are employed by many malware families to make preemptive command-and-control (C&C) countermeasures difficult. Using DGAs, a malware sample may generate hundreds to tens-of-thousands of domain names daily. The domains are generated pseudo-randomly using a seed that is shared by both the malware and the threat actor, so that the threat actor knows *a priori* the sequence of connection attempts by the malware. This represents an asymmetric attack since the defender must sinkhole, pre-register or blacklist *all* of the domains to prevent the C&C connection, while the malware need only connect to a single domain that has been registered by the threat actor.

There are a variety of strategies to detect domain names that are produced by a DGA. Previous works include using a Hidden Markov Model (HMM) framework to model the generating distribution of several DGA families as well as "normal" domains [3], and making bulk predictions on large sets of domains using clustering as a filtering technique [4], [5]. Many methods require contextual information separate from the domain name itself [6]. However, in this paper we restrict our focus to machine learning models that distinguish DGA domains from "normal" domains based solely on the domain name, without contextual information.

This paper explores the use of a generative adversarial network (GAN) to pseudo-randomly produce domain names that are difficult for modern DGA classifiers to detect. The proposed technique generates domains on a character-by-character basis and greatly exceeds the stealth of typical DGA techniques which use simpler algorithms to draw random characters to compose novel domain names. In contrast to simpler DGAs, we propose a DGA architecture that provides a more direct objective: optimize psuedo-random generation of domain names that are by construction difficult for a DGA classifier to detect. This explicit optimization is done using a DGA language model generator coupled with a DGA detector in a GAN [7]. A similar adversarial approach was previously explored by [8] to generate synthetic samples for malware classification based on the DREBIN Android malware dataset. The authors successfully retrained a binary classifier using adversarially crafted input to harden the original classifier.

---

[1] In this paper, we employ a significantly broader meaning of *adversarial examples* as originally used in [2], [1]. In our work, they are not restricted to small perterbations to existing samples, but generally denote artificial samples that "appear realistic" to confound either human or model or both.

In turn, we show increased robustness of a DGA detector to never-before-seen DGA families when trained on an augmented dataset that includes adversarial examples.

Contributions of this paper include the following:

- We present the first known use of a deep learning architecture to pseudo-randomly generate domain names that are by construction difficult for a classifier to distinguish from real domain names.
- We demonstrate that adversarially-generated domain names discovered for a deep learning model are also adversarial to a totally different model architecture (random forest with human-engineered features).
- We demonstrate experimentally that the same adversarial examples can be used to harden the random forest classifier to never-before-seen DGA families.

## II. BACKGROUND

We first provide a brief review in this section of common DGA algorithms, machine learning detection approaches, and follow with background on neural network architectures that we employ in our neural language model for domain generation.

### A. Domain Generation Algorithms

Domain generation algorithms are used by many strains of malware for C&C, including ransomware like `cryptolocker` [9], [10] and `cryptowall` [11], banking trojans, such as `hesperbot` [12], and information stealers, such as `ramnit` [13]. In part, this paper compares DeepDGA to character-based DGA algorithms reproduced from published literature.

Traditional DGA techniques vary in complexity from simple approaches that draw characters uniformly at random, to those that attempt to mimic character or word distributions found in real domains. The `ramnit` DGA, for example, creates domain names using a combination of multiplies, divides and modulos starting from a random seed [13]. On the other hand, `suppobox` creates domains by concatenating two pseudo-randomly chosen English dictionary words [14].

Some example domains from each of the families we consider in this paper are shown in Table I, which are all character-level DGA algorithms. Other common DGAs, like `beebone` have a rigid structure, producing domains like `ns1.backdates13.biz` and `ns1.backdates0.biz`. The `symmi` DGA produces nearly-pronounceable domain names like `hakueshoubar.ddns.net` by drawing a random vowel or a random consonant at each even-numbered index, drawing a random character of the opposite class (vowel/consonant) in the subsequent index location, and appending a second and top-level string like `.ddns.net`.

The unigram distributions for four DGA families and Alexa are shown in Fig. 1. The distributions for `cryptolocker` and `ramnit` are both nearly uniform over the same range. This is expected as they are both generated using a series of multiplies, divisions and modulos based on a single seed [13], [10]. On the other hand, `suppobox` is interesting as it generates unigrams similar to distributions seen by the Alexa top one million domains. The `suppobox` DGA constructs domain names by concatenating multiple randomly chosen words from the English dictionary, and thus follows a similar character distribution to the Alexa top one million. In this paper, we demonstrate a character-based generator composed of a deep learning model that also mimics the distribution of Alexa domain names.

TABLE I
EXAMPLES OF DOMAIN NAMES FROM DGA ALGORITHMS CONSIDERED IN THIS PAPER

| | |
|---|---|
| corebot | ep16g6gjwfixyhs8gfy.ddns.net |
| | ev5texifc43nebil3pk.ddns.net |
| | gf7bm4163fmjkje.ddns.net |
| cryptolocker | agryjvdaabkyt.ru |
| | pwitjnqgjfaqm.org |
| | dhhubfepcdgfv.co.uk |
| dircrypt | hedhryendqlss.com |
| | lgnggnlufbtyjpnvct.com |
| | tzrbdmhoumoy.com |
| kraken_v2 | fwulvdmdytm.com |
| | gybuisybe.cc |
| | gyinkvye.net |
| lockyv2 | btlwubflhfllshn.info |
| | cpgcjsysfwuwa.click |
| | jlbroeji.biz |
| pykspa | gqjgflhop.net |
| | gqumcwaa.org |
| | jpivjh.net |
| qakbot | fgfifkyfut.info |
| | flzuzsaekkipatbtet.biz |
| | owpbsjekk.com |
| ramdo | kugmywaaiymaegiq.org |
| | ocywskaagmmqscoc.org |
| | uomywsaaqggiwouo.org |
| ramnit | byqdmekgd.com |
| | dpmdbwwcmpk.com |
| | gkkcoufektvhiqr.com |
| simda | gatyfusyfi.com |
| | lyvyxoryco.com |
| | puvyxilomo.com |

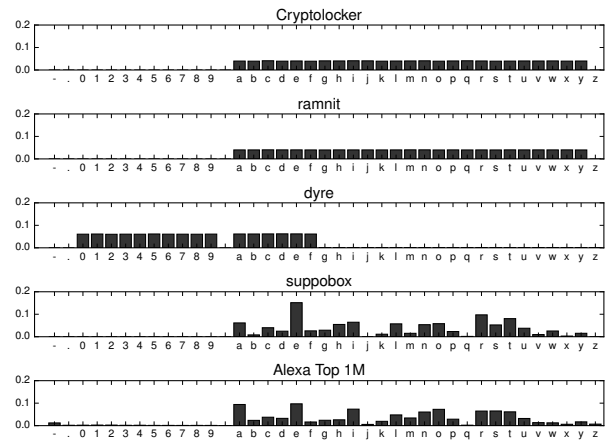

Fig. 1. Unigram distributions for `cryptolocker`, `ramnit`, `dyre` (all three of which are simple character DGAs), `suppobox` (dictionary-based DGA) and the `Alexa` top one million.

*B. DGA detection algorithms*

Previous works in DGA classification approach the problem by either classifying domains in groups to take advantage of bulk statistical properties or common contextual information; or by classifying domains individually with no additional contextual information. We briefly discuss examples of both here, but note that our work falls in the latter category, which may be used in concert with other approaches for DGA detection.

Authors in [4], [5] detect DGAs by using both unigram and bigram statistics of domain clusters. A training set is separated into two subsets: those generated by a DGA and those not generated by a DGA. The distributions of both unigrams and bigrams are calculated for both the subsets. Classification occurs in batches. Each batch of unknown domains (DNS responds with `NXDOMAIN`) is clustered by shared second level domain and domains sharing the same IP address. The unigram and bigram distributions are calculated for each cluster and compared to the two known (labeled) subsets using the Kullback-Leibler (KL) distance. In addition, the authors use the Jaccard distance to compare bigrams between clusters and the known (labeled) sets as well.

Authors in [3] apply a similar clustering process to classify domains with unsuccessful DNS resolutions. To train, statistical features are calculated for each subset of labeled DGA generated domains, such as `Bobax`, `Torpig`, and `Conficker.C`. Unknown domains are clustered by statistical characteristics such as length, entropy, and character frequency distribution as well as shared hosts requesting the domain (i.e., cluster two domains together if the same host made a DNS query for both domains). Next, statistical features are calculated for each cluster and compared to the training subsets to classify the clusters as formed by a known DGA. If a cluster is classified as belonging to a known DGA, the host is deemed to be infected.

Once a host is deemed to be infected with a DGA-bot, the authors attempt to identify the bot's active C2 server. This stage of the process uses a Hidden Markov Model trained on each known family of DGA and applied to single domains (i.e., this technique follows the same assumptions as the LSTM technique proposed by this paper). Each domain with a successful DNS request is fed through each HMM. If a domain receives an adequate score (i.e., greater than some threshold $\theta$), the domain is labeled as a DGA. The threshold is learned at training time and set to a maximum false positive rate of 1%. We use this HMM technique as one of our comparisons to previous work.

Authors in [15] also present a DGA classifier with the intention of classifying individual domains. This classifier uses two basic linguistic features named *meaningful characters ratio* and *n-gram normality score*. The *meaningful characters ratio* calculates the ratio of characters in a domain that comprise of a meaningful word. For example, *endgame* has a ratio of 1 as all characters in the domain are covered by the words *end* and *game* while *game1234* has a ratio of 0.5 as only half of its characters are covered by the word *game*. The *n-gram normality score* is calculated by finding n-grams with $n \in 1, 2, 3$ within a domain and calculating their count in the English language. The mean and covariance of these four features are calculated from a benign set (*Alexa* top 100,000). Unknown domains are then classified by their Mahalanobis distance to the benign set (i.e. a larger distance is indicative of a DGA generated domain). The entire approach is used as a filtering step. Once domains have been classified as a DGA they are fed to a clustering technique (similar to those described above) to further classify the domains.

In our experiments, we leverage a random forest classifier trained on features defined in [3], [4], [5], [15]. We do not implement the full system as defined in [3], [4], [5] as it is based on domain clustering and our intent is to classify DGAs on a per-domain basis. The full system in [15] is not evaluated as they use contextual features such as IP addresses. We assume no contextual information in our experiments, but note that adding contextual information may generally improve a model's ability to detect DGAs.

*C. Adversarial Examples and Generative Adversarial Networks*

Previous work discovered that many machine learning models, including modern neural network architectures, are vulnerable to *adversarial examples*[2], [1]. Notably, [1] introduced the *fast gradient sign method* to systematically discover adversarial examples by perturbing a known "good" sample $\mathbf{x}$ by a small amount $\Delta \mathbf{x} = \epsilon \, \text{sign}\left(\nabla_x J\left(\theta, \mathbf{x}, y\right)\right)$, where $\theta$ represents the model parameters, and $J$ is the cost incurred for classifying $\mathbf{x}$ as class $y$.

Separately, [7] proposed generative adversarial networks as a framework for generating artificial samples that are drawn from the same distribution as the training dataset. Generative adversarial networks incorporate a pair of models—a generator and a discriminator—that compete against each other in a series of adversarial rounds. In the context of our application, the generator learns to create new artificial domain names, and the detector subsequently learns to distinguish the generator's artificial domains from the true domain data distribution.

Previous works apply adversarial examples and GANs to natural images. In this work, we somewhat conflate the use of GANs with the intent of adversarial examples, using a GAN to produce artificial domains and subsequently harden a natural language DGA detector via adversarial training.

*D. Recurrent Neural Network*

In a variety of natural language tasks, recurrent neural networks (RNNs) have been used to capture meaningful temporal relationships among tokens in a sequence [16], [17], [18], [19]. The key benefit of RNNs is that they incorporate contextual (state) information in their mapping from input to output. That is, the output of a single RNN cell is a function of the input layer and previous RNN activations. Due to long chains of operations that are introduced by including self-recurrent connections, the output of a traditional

RNN may decay exponentially over time (or, more rarely but catastrophically explode) for a given input, leading to the well-known *vanishing gradients* problem. This makes learning long-term dependencies in an RNN difficult to achieve.

The problem of vanishing gradients is a key motivation behind the application of the Long Short-Term Memory (LSTM) cell [20], [21], [22], which consists of a state that can be read, written or reset via a set of programmable gates. In the following we consider a layer of LSTM cells using vector notation (boldface), and denote the time index where necessary with subscript $t$. Superscripted $\mathbf{W}$ and $\mathbf{U}$ correspond to particular weight matrices on the input $\mathbf{x}$ or emission $\mathbf{h}$, respectively, and superscripted $\mathbf{b}$ denotes a particular bias vector.

LSTM cells' states $\mathbf{c}$ have self-recurrent connections that allow each cell to retain state between time steps:

$$\mathbf{c}_t = \mathbf{f} \cdot \mathbf{c}_{t-1} + \mathbf{i}_t \cdot \mathbf{g}_t,$$

where · denotes elementwise (Hadamard) multiplication. However, states may be updated in an additive manner by state updates

$$\mathbf{g}_t = \tanh\left(\mathbf{W}^g \mathbf{x}_t + \mathbf{U}^g \mathbf{h}_{t-1} + \mathbf{b}^g\right)$$

via input gates $\mathbf{i}$, which effectively multiply the state update to each cell by a number that ranges between 0 and 1. Likewise, forget gates $\mathbf{f}$ modulate the self-recurrent state connection to each cell's state by a number between 0 and 1. Thus, if the input gate modulates the state update with 0, and the forget gate modulates the recurrent connection with 1, the cell ignores the input and perfectly retains state. On the other hand, a 1 (input) and a 0 (forget) causes a cell's state to be overwritten by the input. And in the case of a 0 (input) and 0 (forget), the state is reset to 0. Finally, output gates $\mathbf{o}$ modulate the contribution of each cell's states to the cell's emission (output) as

$$\mathbf{h}_t = \mathbf{o}_t \cdot \tanh\left(\mathbf{c}_t\right),$$

which propagate to the input gates of LSTM cells across the layer, as well as to subsequent layers of the network. In particular, the input, forget, and output gates are defined as functions of the input $\mathbf{x}_t$ at time $t$ and previous LSTM layer emission $\mathbf{h}_t$ at time $t$, respectively, as

$$\mathbf{i}_t = \sigma\left(\mathbf{W}^i \mathbf{x}_t + \mathbf{U}^i \mathbf{h}_{t-1} + \mathbf{b}^i\right)$$
$$\mathbf{f}_t = \sigma\left(\mathbf{W}^f \mathbf{x}_t + \mathbf{U}^f \mathbf{h}_{t-1} + \mathbf{b}^f\right)$$
$$\mathbf{o}_t = \sigma\left(\mathbf{W}^o \mathbf{x}_t + \mathbf{U}^o \mathbf{h}_{t-1} + \mathbf{b}^o\right).$$

The LSTM cell's design with multiplicative gates allows a network to store and access state over long sequences, thereby mitigating the vanishing gradients problem. For our use with domain names, the state space is intended to capture combinations of tokens that are important to modeling domain names.

### E. Highway Networks

Highway networks were recently proposed as a natural extension of gated memory networks like the LSTM unit to feedforward networks [23]. Highway layers allow for training deep networks by adaptively carrying some dimensions of the input directly to the output through the use of gates. Concretely, the output $\mathbf{y}$ of a single highway layers is the elementwise convex combination of the raw input $\mathbf{x}$ and the transformed input $g(\mathbf{Wx} + \mathbf{b})$ with a vector parameter $\mathbf{t} \in [0,1]^d$:

$$\mathbf{y} = \mathbf{t} \cdot g(\mathbf{Wx} + \mathbf{b}) + (\mathbf{1} - \mathbf{t}) \cdot \mathbf{x},$$

with activation function $g$. In addition to learning the weights $\mathbf{W}$ and bias $\mathbf{b}$, a highway layer also learns the gating parameters $\mathbf{t}$ during training.

## III. METHOD

In this section we describe our DGA neural language architecture and training mechanism. In a first step, we learn to represent valid domain names using an autoencoder architecture, shown in Figure 2(a) and detailed in Section III-A. We then repurpose the encoder (which accepts a domain name and outputs a domain embedding) as a discriminative model, and the decoder (which accepts a domain embedding and outputs a domain name) as a generative model, as show in Figure 2(b) and detailed in Section III-B.

### A. Autoencoder

An autoencoder is a type of data representation model that consists of an encoder that transforms an input to a (usually) lower-dimensional representation, and a decoder which aims to reproduce the original input from the low-dimensional embedding. The character-level encoder shown in Figure 2(a) is loosely inspired by the neural language framework of [24], and the decoder is loosely a mirror image of the encoder.

In what follows, let $\mathbb{V}$ denote the set of lowercase valid domain characters. The encoder contains an embedding layer which learns a linear mapping from $\mathbb{V} \mapsto \mathbb{R}^d$, resulting in a $d$-dimensional vector for each valid domain character. We use $d = 20 < |\mathbb{V}|$ to keep the model size small, and because we don't need to perfectly reproduce the domain characters. Small convolution filters are applied to the name embeddings, which aim to capture simple character combinations present in valid domains. In our implementation, we utilize 20 filters of length 2 (bigrams) and 10 filters of length 3 (trigrams). The next layer selects important features from the convolutional filters via maxpooling and concatenates them into a compact feature vector. Our max pooling actually consists of a *max-over-time* pooling (i.e., max over the symbol sequence) for each of the 30 filters—which measures the presence, but not location, of the bigram and trigram features—as well as a traditional *max-over-filters* pooling—which captures *whether* a bigram/trigram was discovered at a location, but does not preserve *which* bigram/trigram. Assembling the output of maxpooling results in a tensor with 32 dimensions for each time step. This is passed through a highway network (found to improve performance for character-based neural language modeling in [24]), where weights are shared across time-steps, to the input of an LSTM, which accumulates state over the sequence and returns the final emission from the accumulated state as the domain name embedding.

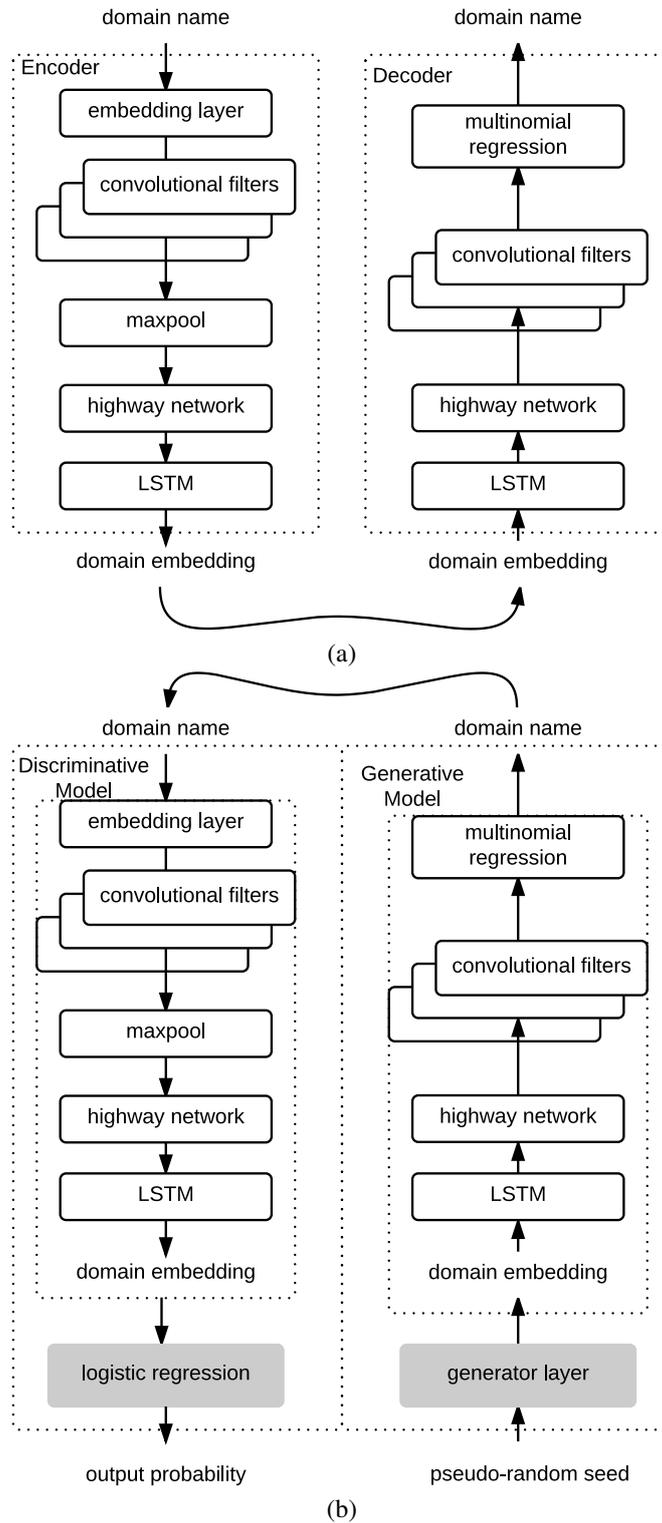

Fig. 2. (a) The domain autoencoder framework is repurposed into a (b) generative adversarial network. Input to the autoencoder in (a) begins in the upper left, while the input to the generative adversarial network begins with the pseudo-random seed in the lower right. During adversarial training, we only train the shaded layers: the generator layer of the generative model and the logistic regression layer of the discriminative model.

The decoder is loosely the reverse of the encoder process. The domain embedding is repeated over the maximum domain name length (time steps), and the resulting sequence is passed as input to an LSTM layer. The sequence of emissions from the LSTM layer are each passed through a highway network with weights shared over time to the same convolution filters as used in the encoder. This results in a 32-dimensional vector for each element in the sequence. The final step is a time-distributed dense layer that acts as a multinomial regressor with weights shared across time steps. Because of a softmax activation on the dense layer, the output of the decoder represents a multinomial distribution over domain characters for each time step, which can be sampled to produce a new domain name that is causally related to the input domain name.

*B. Generative Adversarial Network*

Generative Adversarial Networks, first introduced in [7] for image classification, train two models: A generative model that seeks to create synthetic data based on samples from the true data distribution with added added noise as an input. A discriminator model receives a sample and must predict whether it is a synthetic or a true data sample. This process continues in the form of an adversarial game where the discriminator trains to predict the most accurate label for a sample and the generator trains to construct samples to confound the discriminator.

On its own, the autoencoder described in Section III-A might adequately produce domain names that *look* as if they might be a valid domain (e.g., in the Alexa top 1M), but are actually pseudo-randomly generated by sampling multinomial distributions at the output. However, the use of the autoencoder as a DGA would require that a list of seed domains be stored for use as inputs to the autoencoder. Instead, with only minor modifications to the structure, we repurpose the autoencoder as a GAN that accepts a random seed (number or numbers) as input, and emits a domain name that appears much like a valid domain name.

As shown in Figure 2(b), after the autoencoder has been pretrained on valid domains (e.g., Alexa top 1M), the learned layer weights are frozen. The decoder becomes the key element in a generative model, which merely prepends a dense layer that maps a random input to a domain embedding. Likewise, the encoder becomes a discriminative model, where we simply append a simple logistic regression layer to the domain embedding.

In order to reduce the complexity of the learning task of the generator, we restrict the output space of the generator by a predefined box learned offline from training data. We consider two models: a *box layer* that restricts the output to live in an axis-aligned box defined by embedding vectors of the training data and, a *principal axis box layer* that defines a similar but potentially tighter box, with axes aligned to principal dimensions that represent right singular vectors of the training dataset. Like a traditional sigmoid-activated dense layer, this *box layer* learns a weight matrix $\mathbf{W}$ and bias vector $\mathbf{b}$ to produce a vector $\mathbf{a} = \sigma(\mathbf{W}\mathbf{x} + \mathbf{b})$. However, rather than passing this result $\mathbf{a} \in [0,1]^d$ to the output, we instead use $\mathbf{a}$ as a parameterization of a vector that lives within an axis-aligned box, and pass

$$\mathbf{y} = \mathbf{a} \cdot \mathbf{v}_{\max} + (\mathbf{1} - \mathbf{a}) \mathbf{v}_{\min},$$

as the output, where $\mathbf{v}_{\min}$ and $\mathbf{v}_{\max}$ are, respectively, the minimum and maximum corners of a box in a $d$-dimensional space. These vectors are set *a priori* simply as the elementwise minimum and maximum over all embedded vectors in the training set produced using the first half (encoder) of the autoencoder. The *principal axis box layer* is nearly identical, except that $\mathbf{v}_{\min}$ and $\mathbf{v}_{\max}$ represent the corners of the principal components of the data in the rotated space, and $\mathbf{y}$ is subsequently multiplied by the right singular vector matrix $V$ to transform back into the ambient space. We found in preliminary experiments that the effect of both methods are similar, and for simplicity, leverage the simple box layer in experiments. The use of the box layer as the generator allows the generator to learn embeddings within the domain of embedded vectors without requiring it to learn the extent of the domain. The hope is that the generator need only focus on learning the manifold of Alexa-like domains within the predefined box.

With the weights to the original autoencoder frozen, we train the generator layer and the logistic regression layer by linking the generator and discriminative model together as a GAN. Then, we roughly follow the GAN training procedure introduced in [7], in which the discriminative and generative model compete in adversarial rounds. In our setup the logistic regression weights are trained to separate valid domain names from names produced by the generative model, then the generative model learns weights for its generator layer by targeting an output of 0 (valid domain) for any random input to the combined GAN.

In a slight departure from [7], we regularize the discriminative model by training not only on the most recently-generated samples from the generative model, but a sampled history of domain names from both the current and previous adversarial rounds. This allows the discriminator to "remember" deficiencies in model coverage, and subsequently the generator is forced to learn novel domain embeddings and retreat from a common failure mode of GANs: that without care, samples produced by the generator can collapse to a single point. Subsequent to our experiments, authors in [25] proposed *minibatch discrimination* as another way to prevent this common failure mode. Our approach relies on the regularized discriminator to indirectly prevent the generator from entering the failure mode, while the *minibatch discrimination* encourages diversity of generated samples within a minibatch.

## IV. EXPERIMENTAL SETUP

We implement our architecture in Python using Keras [26]. We train the autoencoder on the Alexa top 1M dataset. Likewise, we train the GAN to distinguish the Alexa top 1M domains from pseudorandomly generated domains generated by the first half (encoder) of the autoencoder. We found that only a few adversarial rounds were required to learn appropriate generator weights when default Keras learning rates (adam optimizer) and batch sizes of 128 were used. For smaller learning rates, more iterations may be required.

We train our generator to accept 20 uniformly distributed numbers using numpy.random.rand, which allows for a seed, in order to produce a fake domain name. Similarly, we use numpy.random.multinomial to sample from the output using the common seed.

Although the GAN framework ensures that DeepDGA maximally confuses the detector model, we are interested in its ability to bypass an independent model. So, to measure detectability of the DGA, we measure detection rates using a random forest DGA classifier that uses manually-crafted domain name features defined in [3], [4], [5], [15]. In particular, the manually crafted features of the random forest DGA classifier include the following:

- length of domain name,
- entropy of character distribution in domain name,
- vowel to consonant ratio,
- Alexa top 1M $n$-gram frequency distribution co-occurrence count, where $n = 3, 4$ or $5$,
- *n-gram normality score*, and
- *meaningful characters ratio*.

Note that for the $n$-gram normality score, we use $n = 3$, $n = 4$ and $n = 5$ as three distinct features as opposed to $n = 1$, $n = 2$ and $n = 3$ as in [15] since the larger $n$-gram size performed better in preliminary experiments.

## V. RESULTS

The autoencoder was pretrained for 300 epochs, with each epoch using 256K domains randomly sampled from the Alexa Top 1M, and a batch size of 128. Pretraining required roughly 14 hours on a single NVIDIA Titan X GPU. Subsequent to pretraining, in each adversarial round, we generated 12,800 adversarial samples against the detector. Each round required roughly 7 minutes on the GPU. Results from the training are detailed in the following subsections.

### A. Autoencoder results

Given a domain as input, the autoencoder produces a multinomial distribution over the possible characters (outcomes), from which a domain can be sampled. For example, a few domains sampled from the output of the autoencoder given the input (with TLD removed) are shown in Table II.

The autoencoder does not reconstruct perfectly the input, which can be ascribed to the stochastic sampling of the multinomial distributions to choose each domain character via independent draws, and insufficient model capacity (e.g., heavy model bias) to express all character combinations of

TABLE II
A FEW DOMAINS SAMPLED FROM THE OUTPUT OF THE DEEPDGA
AUTOENCODER (WITH TLDS REMOVED)

```
    biogartenversand --> bidegerafernrnps
             metapack --> mdtkhrhk
          diaboliclabs --> dirinoyslhns
  homehealthsollution --> tsntneyerasyeeaenne
           mousebreath --> metbrslamtl
         newinventions --> nataiirsfsdos
             wonderbox --> weonkimwb
            tara-china --> tun-mos-gi
                gobuu --> caxfx
              keeyword --> kaxioawy
              kandidco --> kauircae
     toaster-schwerin --> teztoylabr-bs-st
```

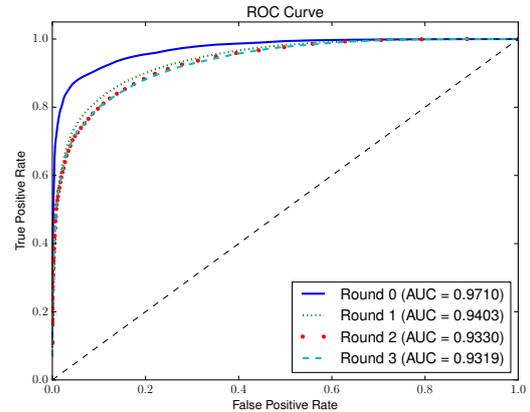

Fig. 3. ROC curves for random forest classifier trained on DeepDGA vs. Alexa top 10K averaged over 10-fold crossvalidation. DeepDGA becomes increasingly more difficult for the random forest classifier to detect with each round of adversarial training up to three rounds. (Subsequent rounds don't produce substantial AUC reduction.)

domains in the Alexa top 1M. However, since we actually don't want to generate domain names in the Alexa top 1M, but rather names names from the same (or similar) generative distribution, the reconstructions are wholly adequate.

### B. GAN results

Figure 3 displays Receiver Operating Characteristic (ROC) curves of a random forest classifier after each of four adversarial rounds. The classifier is trained on DeepDGA generated domains as malicious and the Alexa top 10K as benign. Results are based on 10-fold cross-validation. The ROC curves demonstrate that performance of the random forest classifier degrades with the number of adversarial rounds with an apparent asymptote after three rounds. This degradation is due to the GAN's ability to generate domains that create confusion for the classifier. Training on these domains will allow us to harden the classifier to blind spots to increase performance of a DGA detector. All subsequent experiments in this paper will be based on three adversarial rounds due to decreasing utility after three rounds.

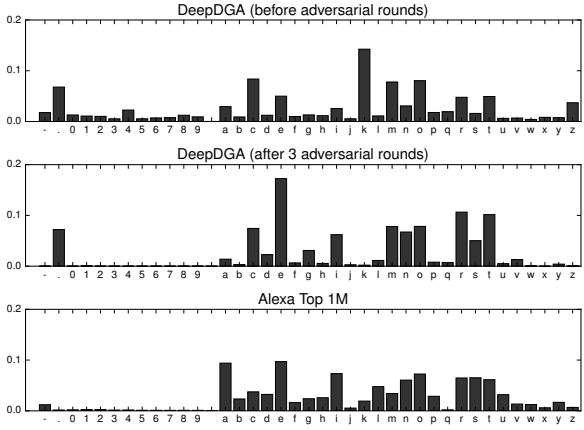

Fig. 4. Character distribution of DeepDGA compared to the Alexa top 1M before adversarial rounds and after 3 adversarial rounds.

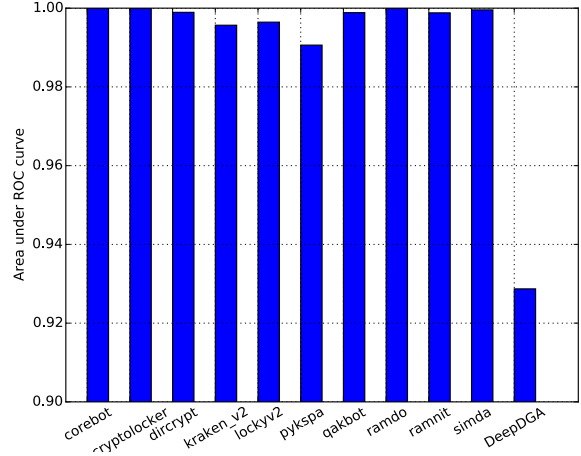

Fig. 5. Area under ROC curve for random forest classifier trained on each DGA algorithm vs. Alexa top 10K averaged over 10-fold crossvalidation.

Figure 4 displays the unigram distribution of domains generated before adversarial rounds, after three adversarial rounds, and those from the Alexa top 1M. Note that the unigram distribution approaches that of the Alexa top 1M after three rounds demonstrating the confusing nature of domains generated by DeepDGA.

Compared to other character-based DGAs, DeepDGA shows significant improvement in its ability to produce domains that are undetectable by a DGA classifier. Figure 5 shows the area under the ROC curve for the random forest classifier trained to detect each of ten character-based DGAs and DeepDGA vs. the Alexa top 10K *individually*. In other words, there are eleven different random forest models, each trained *specifically* to detect a particular DGA. Using the the random forest model, DeepDGA exhibits a false negative rate of roughly 1 in 14, as opposed to the next-best-performaing DGA, `pykspa`, which exhibits a false negative rate of roughly 1 in 106, a decrease of over 7×.

The detection rate is even more striking when one trains a general DGA detector on the available samples. We create a dataset using the top 10K Alexa domain names and 10K domain names from each of the DGAs in the comparison set, including DeepDGA. We report the average crossvalidation score over 10-folds using a 20% holdout set for each fold. Results in Table III show that under such a scenario, less than 50% of the DeepDGA samples are detected by the classifier (recall), whereas the next best DGAs (`simda` and `kraken_v2`) are detected at a rate of 98%. This represents a 25× improvement in avoiding detection (1 in 2 vs. 1 in 50).

A few DeepDGA domains (without TLDs) after adversarial tuning are shown below in Table IV. Domain names are of varying length, exhibit no strong common patterns, and can be exactly reproduced via a random seed.

### C. Hardening a machine learning model

It has previously been shown that adversarial examples may be shared across different machine learning models [2],

TABLE III
DETECTION RATE ON 20% HOLDOUT SETS AVERAGED OVER 10 CROSSVALIDATION FOLDS OF TRAINING THE RANDOM FOREST CLASSIFIER, WITH EQUAL NUMBER OF SAMPLES PER CLASS.

|             | detection rate | support |
|---|---|---|
| corebot     | 1.00 | 1997 |
| cryptolocker | 1.00 | 1988 |
| dircrypt    | 0.99 | 1977 |
| kraken_v2   | 0.96 | 1998 |
| lockyv2     | 0.97 | 1982 |
| pykspa      | 0.85 | 2009 |
| qakbot      | 0.99 | 1985 |
| ramdo       | 0.99 | 1985 |
| ramnit      | 0.98 | 2015 |
| simda       | 0.96 | 2025 |
| DeepDGA     | **0.48** | 2497 |
| avg / total | 0.96 | 22457 |

[1]. We demonstrate that by augmenting a training set with adversarial examples generated by the GAN, a model can be hardened against DGA families not observed in the training set. In particular, we trained the random forest model using a leave-one-family-out strategy in which an entire DGA family is held out for validation, while the random forest model is trained on the remaining nine families. The top Alexa 10K are included in the training set, and the next Alexa 10K are included in the holdout set. This baseline result is compared to a hardened result in which 10K DeepDGA samples are appended to the 9-family plus Alexa training dataset. For each case, we report the true positive rate (TPR) at a fixed false positive rate (FPR) of 1%. This generally required different thresholds for baseline and hardened models.

TABLE IV
DOMAINS NAMES (WITHOUT TLDS) AFTER ADVERSARIAL TUNING

```
firiaps     sirgivrv    laner         mivognit
qiurdeees   tisehl      spienienitne  yhujq
gyldles     thellehm    thuaemoa      statpottxy
lirneret    chdareet    shtrunoa      vietips
```

TABLE V
BINARY CLASSIFICATION OF A RANDOM FOREST CLASSIFIER BEFORE AND AFTER ADVERSARIAL HARDENING. EACH ROW REPRESENTS THE TRUE POSITIVE RATE (TPR) AT A FIXED 1% FALSE POSITIVE RATE ON A HELD-OUT DGA FAMILY OF 10K DOMAINS AND 10K HELD-OUT ALEXA DOMAINS. ALL OTHERS FAMILIES OF 10K SAMPLES (PLUS ALEXA TOP 10K) ARE USED IN THE TRAINING SET, WITH 10K ADVERSARIAL EXAMPLES BEING AUGMENTED IN THE CASE OF THE HARDENED CLASSIFIER.

|  | baseline | hardened |
|---|---|---|
| `corebot` | **0.97** | **0.97** |
| `dircrypt` | **0.95** | 0.93 |
| `qakbot` | **0.94** | **0.94** |
| `ramnit` | **0.94** | **0.94** |
| `lockyv2` | **0.87** | 0.84 |
| `cryptolocker` | 0.87 | **0.88** |
| `simda` | 0.75 | **0.79** |
| `krakenv2` | 0.72 | **0.76** |
| `pykspa` | 0.67 | **0.71** |
| `ramdo` | **0.54** | **0.54** |
| average | 0.68 | **0.70** |

Training the classifier on adversarially crafted samples generally improved the model's ability to detect families not in the training set. Table V shows that the hardened classifier maintains or increases the effective TPR especially for families with low baseline TPRs, and all families except `dircrypt` and `lockyv2` which exhibit marginally smaller effective TPRs at a 1% FPR.

## VI. DISCUSSION

We have demonstrated automatically generating artificial domains that are adversarial to a deep learning DGA model. The adversarial examples are shown to be shared between the deep learning detector model—for which they were explicitly optimized to circumvent—as well as a random forest model. This demonstrates in an information security setting a key point in [2], [1]: that adversarial examples may be shared across different models.

Training a GAN to generate these examples requires substantial art to prevent common failure modes. We introduced novel history regularization, neural layers (*box layer* and *principal axis box layer*), and more common autoencoder pre-training to simplify the learning task of the generator. Subsequent to our experiments, [25] proposed other strategies for training GANs, which we leave to future work.

Furthermore, we have demonstrated that by augmenting a training set with DeepDGA adversarial examples, a random forest classifier was hardened against DGA families not observed during training. The ability to harden using GAN-crafted samples generally increased TPR for a fixed FPR in our experiments. Unlike the perterbation-based adversarial example generation proposed in [1] (i.e., *fast gradient sign method*), the GAN-crafted samples are meant to match the data actual distribution, so that, without care, FPR may be adversely affected when used for hardening. A qualitative comparison of the adversarial example quality and quantitative comparison of hardening strategies using the more direct fast-gradient sign method for DGA detection is left to future work.

## REFERENCES

[1] I. J. Goodfellow, J. Shlens, and C. Szegedy, "Explaining and harnessing adversarial examples," *arXiv preprint arXiv:1412.6572*, 2014.
[2] C. Szegedy, W. Zaremba, I. Sutskever, J. Bruna, D. Erhan, I. Goodfellow, and R. Fergus, "Intriguing properties of neural networks," *arXiv preprint arXiv:1312.6199*, 2013.
[3] M. Antonakakis, R. Perdisci, Y. Nadji, N. Vasiloglou, S. Abu-Nimeh, W. Lee, and D. Dagon, "From throw-away traffic to bots: detecting the rise of DGA-based malware," in *P21st USENIX Security Symposium (USENIX Security 12)*, pp. 491–506, 2012.
[4] S. Yadav, A. K. K. Reddy, A. Reddy, and S. Ranjan, "Detecting algorithmically generated malicious domain names," in *Proc. 10th ACM SIGCOMM conference on Internet measurement*, pp. 48–61, ACM, 2010.
[5] S. Yadav, A. K. K. Reddy, A. N. Reddy, and S. Ranjan, "Detecting algorithmically generated domain-flux attacks with DNS traffic analysis," *Networking, IEEE/ACM Transactions on*, vol. 20, no. 5, pp. 1663–1677, 2012.
[6] A. J. Aviv and A. Haeberlen, "Challenges in experimenting with botnet detection systems.," in *CSET*, 2011.
[7] I. Goodfellow, J. Pouget-Abadie, M. Mirza, B. Xu, D. Warde-Farley, S. Ozair, A. Courville, and Y. Bengio, "Generative adversarial nets," in *Advances in Neural Information Processing Systems*, pp. 2672–2680, 2014.
[8] N. Papernot, P. McDaniel, X. Wu, S. Jha, and A. Swami, "Distillation as a defense to adversarial perturbations against deep neural networks," in *Proceedings of the 37th IEEE Symposium on Security and Privacy*, 2015.
[9] M. Ward, "Cryptolocker victims to get files back for free," *BBC News, August*, vol. 6, 2014.
[10] "A closer look at cyrptolocker's DGA." https://blog.fortinet.com/post/a-closer-look-at-cryptolocker-s-dga. Accessed: 2016-04-22.
[11] N. Hampton and Z. A. Baig, "Ransomware: Emergence of the cyber-extortion menace," in *Australian Information Security Management Conference*, 2015.
[12] A. Cherepanov and R. Lipovsky, "Hesperbot-A new, advanced banking trojan in the wild," 2013.
[13] Symantec, *W32.Ramnit analysis*. 2015-02-24, Version 1.0.
[14] J. Geffner, "End-to-end analysis of a domain generating algorithm malware family." Black Hat USA 2013, 2013.
[15] S. Schiavoni, F. Maggi, L. Cavallaro, and S. Zanero, "Phoenix: DGA-based botnet tracking and intelligence," in *Detection of intrusions and malware, and vulnerability assessment*, pp. 192–211, Springer, 2014.
[16] A. J. Robinson, "An application of recurrent nets to phone probability estimation," *Neural Networks, IEEE Transactions on*, vol. 5, no. 2, pp. 298–305, 1994.
[17] T. Mikolov, M. Karafiát, L. Burget, J. Cernockỳ, and S. Khudanpur, "Recurrent neural network based language model.," in *INTERSPEECH*, vol. 2, p. 3, 2010.
[18] A. Graves, "Sequence transduction with recurrent neural networks," *arXiv preprint arXiv:1211.3711*, 2012.
[19] Y. Bengio, N. Boulanger-Lewandowski, and R. Pascanu, "Advances in optimizing recurrent networks," in *Acoustics, Speech and Signal Processing (ICASSP), 2013 IEEE International Conference on*, pp. 8624–8628, IEEE, 2013.
[20] S. Hochreiter and J. Schmidhuber, "Long short-term memory," *Neural computation*, vol. 9, no. 8, pp. 1735–1780, 1997.
[21] F. A. Gers, J. Schmidhuber, and F. Cummins, "Learning to forget: Continual prediction with LSTM," *Neural computation*, vol. 12, no. 10, pp. 2451–2471, 2000.
[22] F. A. Gers, N. N. Schraudolph, and J. Schmidhuber, "Learning precise timing with LSTM recurrent networks," *J. Machine Learning Research*, vol. 3, pp. 115–143, 2003.
[23] R. K. Srivastava, K. Greff, and J. Schmidhuber, "Highway networks," *arXiv preprint arXiv:1505.00387*, 2015.
[24] Y. Kim, Y. Jernite, D. Sontag, and A. M. Rush, "Character-aware neural language models," *arXiv preprint arXiv:1508.06615*, 2015.
[25] T. Salimans, I. Goodfellow, W. Zaremba, V. Cheung, A. Radford, and X. Chen, "Improved techniques for training gans," *arXiv preprint arXiv:1606.03498*, 2016.
[26] F. Chollet, "keras." https://github.com/fchollet/keras, 2016.